\documentclass[a4paper,twocolumn,showpacs,superscriptaddress,floatfix,nofootinbib,prd]{revtex4-1}
\usepackage{graphicx}
\usepackage{epsf}
\usepackage{epsfig}
\usepackage{amssymb,amsmath}
\usepackage[usenames]{color}
\usepackage{amssymb}
\usepackage{wasysym}
\usepackage{soul}
\usepackage{times}
\usepackage{mathrsfs}
\usepackage{hyperref}
\usepackage{booktabs}
\usepackage{chngcntr}
\hypersetup{
  colorlinks=true,          linkcolor=blue,           citecolor=cyan,         }
\usepackage{float}
\usepackage{tensor}

\definecolor{orange}{rgb}{1,0.5,0}

\def\prd{{\it Phys. Rev.} D~}

\def\apjl{{\it Astrophys. J.} Lett~}
\def\apj{{\it Astrophys. J.}}
\def\Msun{M_\odot}

\def\mnras{{\it MNRAS~}}

    \def\eg{{\it e.g.}}

\def\lesssim{\mathrel{\hbox{\rlap{\hbox{\lower4pt\hbox{$\sim$}}}\hbox{$<$}}}}
\def\gtrsim{\mathrel{\hbox{\rlap{\hbox{\lower4pt\hbox{$\sim$}}}\hbox{$>$}}}}
\def\alt{\mathrel{\hbox{\rlap{\hbox{\lower4pt\hbox{$\sim$}}}\hbox{$<$}}}}
\def\agt{\mathrel{\hbox{\rlap{\hbox{\lower4pt\hbox{$\sim$}}}\hbox{$>$}}}}

\def\araa{{\it ARA\&A}}

\usepackage{mathtools}

\def\gta{\ifmmode {\mathbin{\lower 3pt\hbox       {$\,\rlap{\raise 5pt\hbox{$\char'076$}}\mathchar"7218\,$}}}
    \else {${\mathbin{\lower 3pt\hbox
    {$\rlap{\raise 5pt\hbox{$\char'076$}}\mathchar"7218\,$}}}
    $}\fi}
\def\lta{\ifmmode {\,\mathbin{\lower 3pt\hbox       {$\,\rlap{\raise 5pt\hbox{$\char'074$}}\mathchar"7218\,$}}}
    \else {${\mathbin{\lower 3pt\hbox
    {$\rlap{\raise 5pt\hbox{$\char'074$}}\mathchar"7218\,$}}}
    $}\fi}

\newcommand{\beq}{\begin{equation}}
\newcommand{\eeq}{\end{equation}}
\newcommand{\bea}{\begin{eqnarray}}
\newcommand{\eea}{\end{eqnarray}}

\renewcommand{\BibitemShut}[1]{}
\newcommand{\ms}{\,{\rm ms}}

\newcommand{\Msol}{M_{\odot}}

\graphicspath{{figs/}}
\begin{document}

\title[f-modes ns structure]{Probing neutron star structure via f-mode oscillations and damping in dynamical spacetime models}

\author{Shawn Rosofsky}
\affiliation{NCSA, University of Illinois at Urbana-Champaign, Urbana, Illinois 61801, USA}
\affiliation{Department of Physics, University of Illinois at Urbana-Champaign, Urbana, Illinois 61801, USA}
\author{Roman~Gold}
\affiliation{Institut f{\"u}r Theoretische Physik,
Johann Wolfgang Goethe-Universit\"at, Max-von-Laue-Stra{\ss}e 1, 60438 Frankfurt,
Germany}
\affiliation{Department of Physics and Joint Space-Science Institute, University of Maryland, College Park, MD 20742-2421, USA}
\affiliation{Perimeter Institute for Theoretical Physics, 31 Caroline Street North, Waterloo, ON, N2L 2Y5, Canada}

\author{Cecilia Chirenti}
\affiliation{Center for Mathematics, Computation and Cognition, UFABC,
Santo Andr\'e - SP, 09210-580, Brazil}
\author{E. A. Huerta}
\affiliation{NCSA, University of Illinois at Urbana-Champaign, Urbana, Illinois 61801, USA}
\affiliation{Department of Astronomy, University of Illinois at Urbana-Champaign, Urbana, Illinois 61801, USA}
\author{M.~Coleman Miller}
\affiliation{Department of Astronomy and Joint Space-Science Institute, University of Maryland, College Park, MD 20742-2421, USA}

\begin{abstract} 
\noindent 
Gravitational wave and electromagnetic observations can provide new insights
into the nature of matter at supra-nuclear densities inside neutron stars.
Improvements in electromagnetic and gravitational wave sensing instruments
 continue to enhance the accuracy with which they
can measure the masses, radii, and tidal deformability of neutron stars.  
These better measurements place tighter constraints on the equation of
state of cold matter above nuclear density.  
In this article, we discuss a complementary approach to get
insights into the structure of neutron stars by providing a model
prediction for non-linear fundamental eigenmodes (f-modes) and their
decay over time, which are thought to be induced by time-dependent tides
in neutron star binaries. Building on pioneering studies that relate the
properties of f-modes to the structure of neutron stars, we
systematically study this link in the non-perturbative regime using
models that utilize numerical relativity. Using a suite of fully
relativistic numerical relativity simulations of oscillating TOV stars,
we establish blueprints for the numerical accuracy needed to accurately
compute the frequency and damping times of f-mode oscillations, which we
expect to be a good guide for the requirements in the binary case. We
show that the resulting f-mode frequencies match established results from
linear perturbation theory, but the damping times within numerical
errors depart from linear predictions. This work lays the foundation for
upcoming studies aimed at a comparison of theoretical models of f-mode
signatures in gravitational waves, and their uncertainties with actual
gravitational wave data, searching for neutron star binaries on highly
eccentric orbits, and probing neutron star structure at high densities.
\end{abstract}

\maketitle

\section{Introduction}
\label{sec:intro}

Neutron stars~\cite{NSLandau} present unique opportunities to 
study the nature of matter at supra-nuclear 
densities at relatively low temperatures, and in 
the presence of extreme gravitational interactions. Given the diverse scenarios 
that have been invoked to describe the
properties of cold, dense matter in the cores of neutron stars (nucleons,
hyperons or free
quarks)~\cite{Stone2016EPJA,Alcock1986ApJ,YakARAAY,Weber2005PrPNP}, it is
essential to confront these theories with observations in the
gravitational or electromagnetic spectrum, or a combination of them
through multimessenger astrophysics discovery
campaigns~\cite{bnsdet:2017,mma:2017arXiv}. These studies will constrain the
space of permissible models and hopefully, gain a better understanding of
the physics at supranuclear densities.  Electromagnetic observations of
the pulsars PSR J1614-2230~\cite{2016Fonseca} and PSR
0348+0432\cite{Anto2013Sci}, which are consistent with neutron stars with
masses \(M\sim 2\Msun\), have already significantly reduced the existing
number of astrophysically viable equations of state
(EOS), \(P(\epsilon)\), i.e., the relation between the pressure \(P\) and
the total energy density \(\epsilon\). In \cite{Rezzolla2018} an upper
limit for the neutron star maximum mass of $M=2.16M_\odot$ was derived
based on GW170817 and basic arguments on kilonovas. A consistent limit was derived from different considerations \cite{2017ApJ...850L..19M} base on previous arguments about the nature of short gamma-ray bursts \cite{2013PhRvL.111m1101B,2015ApJ...812...24F,2015ApJ...808..186L}.

X-ray observations of the pulse profiles generated by hot spots on 
moderately spinning neutron stars with the ongoing NICER~\cite{Gendreau2012SPIE} and
future LOFT~\cite{2012LOFT} missions  may enable measurements of the 
neutron star radii at the 5\% level~\cite{psaltis2014ApJ}, improving existing measurements that are
dominated by systematic errors~\cite{mill2013,millamnb2016EPJA}. 
This precision can be achieved through 
long exposure times using NICER, or using the large collecting area of LOFT. 

LIGO~\cite{DII:2016,LSC:2015} and Virgo~\cite{Virgo:2015} gravitational wave (GW) observations of the 
neutron star merger GW170817~\cite{bnsdet:2017,mma:2017arXiv,2017Sci...358.1556C} imply that the two neutron stars had radii 
\(R_{\{1,\,2\}}=\tensor*[]{\mathrm{11.9}}{^{+1.4}_{-1.4}}\,\textrm{km}\), 
and to place constraints on the EOS at supra-nuclear densities 
\(P(\epsilon)=\tensor*[]{\mathrm{3.5}}{^{+2.7}_{-1.7}}\times 10^{34}\,\textrm{dyn}\,\textrm{cm}^{-2}\) (quantified at the 90\% confidence level~\cite{2018arXiv180511581T}) as well as the dimensionless tidal deformability \(\tilde{\Lambda} \lesssim 800\)~\cite{bnsdet:2017}.  
As a reference point, the GW measurement of \(P(\epsilon)\) for GW170817
is at twice nuclear saturation density, whereas terrestrial laboratories
can only test and constrain cold EOSs at densities near or below
saturation density of nuclei. GW observations enable these measurements
because the waveform models used to analyze LIGO and Virgo data encode
information about the static and dynamic tides that determine the
dynamical evolution of the system, driving it to merger faster than a
system of two point
masses~\cite{Blanchet:2006,Rosswog:2003,2016PhRvD..93l4062H,Parisi2017}.

GW and electromagnetic observations will continue to provide new and
detailed information about the nature of neutron star matter. GW
observations will shed light on the masses and tidal deformability of
neutron stars, while electromagnetic observations will enable accurate
measurements of their mass and radius~\cite{Gendreau2012SPIE} and may
even be informative for studies of stellar structure in the context of
non-GR spacetimes \cite{Mendes2018}.

Even though multimessenger observations will provide new and detailed information 
about the properties of neutron stars, they are likely to leave open questions, especially in light of
potential emission-model dependence, which calls for independent and
complementary measurements. In this article, we will focus on a
different observational scenario to obtain this information from direct
observations of the stars, namely, the measurement of fundamental
eigenmodes (f-modes) of the individual stars, which can be caused by
tidal effects, and for which pressure is the restoring force. 

Proposed scenarios for the generation of f-modes in neutron stars include
neutron star binaries or neutron star-black hole systems in highly
eccentric orbits \cite{Gold2012,East2012b0,East2012c,East2016a,East2016,Papenfort2018}, i.e. binary separations comparable to the radii of the
binary components, such that violent changes in the tidal field can
perturb the stars without necessarily causing an immediate
collision~\cite{chi2017ApJ,Huerta:2017a,Huerta:2014,Huerta:2013a,huerta:2018PhRvD,Rebei2018R}. These
close interactions will induce f-mode oscillations in cold, slowly
rotating neutron stars, thereby simplifying the analysis and modeling of
f-modes as compared to the differentially rotating
scenario~\cite{chi2017ApJ} or shock heated matter in hypermassive neutron
stars (HMNSs) \cite{Bauswein2014,Takami2014}.

Extracting physics encoded in the f-modes from observational data
involves an accurate model prediction \cite{Steinhoff2016,Yang2018},
and measurement of their frequencies \(\omega\) and damping
times \(\tau\). Seminal studies have established that these two
observables are related to the average density and
compactness~\cite{An1998MNRAS,Banhar2004PhRvD}. In \cite{Tsui2005,Lau2010}
universal relations using the effective compactness \(\eta
= \left(M/I^3\right)^{1/2}\) with mass \(M\) and moment of inertia \(I\)
were discovered, and were tested in \cite{Chirenti2015}.

In principle, once \(\omega\) and \(\tau\) are extracted from GW
observations, it may be possible to infer \(M\) and \(I\). Furthermore,
the amplitude of the f-mode depends on the orbital energy deposited in
the mode, and this can be used to measure the Love number \(\lambda\) 
and the rotationally-induced quadrupole \(Q\) through \(I\)-Love-\(Q\)
relations~\cite{flanagan2008PhRvD,K2013Sci,Yagi2013PhRvD,YaYu2014PhRvD}.

In this article, we carry out a detailed analyses to quantify the
required accuracy in numerical relativity simulations to extract the
f-mode oscillations and damping times. To do this, we simulate
oscillating neutron stars using the open source \texttt{Einstein Toolkit}
numerical relativity software, and compare these results to linear
perturbation theory.  This work lays the foundation for future numerical
relativity simulations of highly eccentric neutron star mergers (see~\cite{Dietrich2018} 
for a state-of-the-art library) with the aim of
accurately extracting f-mode oscillation damping times and frequencies as
well as a systematic investigation for a variety of equations of state and the effect of model uncertainties
on parameter estimation.  A successful detection and accurate
parameter estimation from eccentric neutron star binaries would provide a
wealth of information on astrophysical evolution
scenarios \cite{Gondan2018}.

This article is organized as follows: Section~\ref{sec:methods} outlines
the approach we have followed to couple Einstein's field equations with a
relativistic perfect fluid. We describe the construction of initial data
to excite f-mode oscillations in Tolman-Oppenheimer-Volkoff (TOV) stars,
and the extraction of the corresponding GWs. In
Section~\ref{sec:results}, we show that the numerical formalism
introduced in this study exhibits the necessary convergence to accurately
compute the frequency and damping time of f-mode oscillations. We also
demonstrate that our simulations reproduce the expected results from
linear perturbation theory in the appropriate limit. We summarize our
findings and future directions of work in Section~\ref{sec:conclusions}.

\section{Methods}
\label{sec:methods}

We evolve the spacetime by solving the Einstein field equations using
the tools of numerical relativity.

\begin{equation}
R_{\mu\nu}-\frac{1}{2}R g_{\mu\nu} = 8\pi T_{\mu\nu}
\label{eq:EFE}
\end{equation}

We express the four-dimensional metric \(g_{\mu\nu}\) in the standard
3+1 split of spacetime \cite{Alcubierre:2008, Bona2009, Baumgarte2010,
  Gourgoulhon2012, Rezzolla_book:2013, Shibata_book:2016} as

\begin{equation}
  ds^{2} = -\alpha^{2} dt^{2} + \gamma_{ij} (dx^{i} + \beta^{i} dt)(dx^{j} + \beta^{j} dt)\,
  \label{eq:3p1_metric}
\end{equation}
with the spatial metric \(\gamma_{ij}\) induced on each hypersurface of
the spacetime foliation, the lapse \(\alpha\), shift
\(\beta^{i}\), and the hypersurface unit normal vector
\begin{equation}
  n^{\mu}=\left(\alpha^{-1},-\frac{\beta^{i}}{\alpha}\right)
\end{equation}
Projecting the Einstein equations onto the hypersurface and along
\(n^\mu\) yields a set of constraint and evolution equations that
together with suitable initial data form our initial value problem.

\subsection{Spacetime Evolution}
\label{sec:spactime_evolution}

There are infinitely many formulations of the Einstein equations that
are equivalent in the continuum limit of infinite spatial and temporal
resolution, but that have distinct principal parts and hyperbolicity
properties, and hence behave differently at finite resolution.  In this
study we primarily employed the BSSN formalism \cite{Shibata95,Baumgarte99}.

The CCZ4 formalism \cite{Alic:2011a} was also used for some results. The starting
point is the covariant extension of the Einstein equations

\begin{eqnarray}
8\pi\left(T_{\mu\nu}-\frac{1}{2}g_{\mu\nu}T\right) & = & R_{\mu\nu} + 2\nabla_{(\mu}Z_{\nu)} \label{eq:CCZ4} \\
 & + & \kappa_{1}\left[2n_{(\mu}Z_{\nu)}-\left(1+\kappa_{2}\right)g_{\mu\nu}n_{\sigma}Z^{\sigma}\right] \nonumber
\end{eqnarray}
also known as the Z4 formulation, with the four-vector \(Z_{\mu}\)
containing the four constraints. These constraints have to be satisfied 
for all times and in particular
by the initial data used as input for the time integration. If

\begin{equation}
  Z_{\mu} = 0\,
\label{eq:Z0}
\end{equation}
then the Einstein field equations are recovered. This equivalence in the
continuum limit implies that solutions to Eq.~\eqref{eq:CCZ4} under the constraint
\eqref{eq:Z0} are also solutions to Eq.~\eqref{eq:EFE}.  The additional
terms involving the constraint damping coefficients \(\kappa_{1,2}\) lead
to an exponential damping of the constraint violations on a
characteristic timescale proportional to the coefficients. Even in the
zero damping case \(\kappa_1=\kappa_2=0\), Z4-based formulations feature
a \emph{propagating} mode associated with the Hamiltonian constraint
violations as opposed to the classic BSSN formulation
\cite{Shibata95,Baumgarte99}, where a \emph{0-speed} mode leads to local
build-up of Hamiltonian constraint violations due to numerical error. In
Z4-based formulations such violations propagate and therefore have a
chance to leave the grid. It is therefore expected that Z4-based
formulations yield more accurate spacetime evolutions than the BSSN
system.  Employing a conformal and traceless decomposition of the
resulting equations leads to the CCZ4 system, see \cite{Alic:2011a} for
details and \cite{Hilditch2012,Bernuzzi:2009ex} for an alternative
approach.  For the cases in this work in which the CCZ4 formulation was
used, we set \(\kappa_1=0.02\), \(\kappa_{2}=0\).

With that said, the CCZ4 formulation is sensitive to the boundary
conditions and may lead to instabilities if not applied properly
\cite{Alic:2011a}. To ensure robustness for our work, 
BSSN was employed for most of this work.

The evolution of the gauge conditions, i.e., lapse and shift, is
governed by standard 1+log slicing condition and a Gamma driver (see
\eg \cite{Baumgarte2010,Baiotti2016}).

We are making use of the publicly available \verb+McLachlan+ code
\cite{Brown2007b} which implements the above evolution equations with
fourth order central finite differences and is part of the
\verb+Einstein Toolkit+ \cite{Loffler:2011ay}.

The non-linear stability of smoothness properties of the spacetime
evolution is enhanced by adding an artificial Kreiss-Oliger
dissipation \cite{Kreiss73} to the spacetime variables.

\subsection{General Relativistic Hydrodynamics}
\label{sec:grhydro}
The neutron stars are modeled by a relativistic perfect fluid
\begin{equation}
  T_{\mu\nu} = \rho h u_{\mu} u_{\nu} + p g_{\mu\nu}
\end{equation}
where \(u^{\mu}\) is the 4-velocity, \(\rho\) is the rest mass, \(h\) is the
specific enthalpy and \(p\) is the pressure of the fluid.

The standard relativistic hydrodynamic equations are the
local conservation law for the energy-momentum tensor
\begin{equation}
  \nabla_{\mu}T^{\mu\nu} = 0,
  \label{eq:em-conservation}
\end{equation}
and the conservation of rest mass
\begin{equation}
  \nabla_{\mu}\left(\rho u^{\mu}\right) = 0
  \label{eq:rm-conservation}
\end{equation}

Both equations together with the spacetime metric \(g_{\mu\nu}\)
represent the equations of motion of the fluid, which are closed by a
suitable equation of state (EOS) with the general form \(p =
p\left(\rho,T,Y_{e}\right)\) where $T$ is the temperature and $Y_e$ the
so-called electron fraction (low $Y_e$ indicates neutron-rich
material). In this pilot study we restrict ourselfs to polytropes, but in
the future we intend to investigate more realistic equations of state.

These equations can be cast into conservative form, which is also
known as the ``Valencia formulation'' \cite{Banyuls97}

\begin{equation}
  \partial_t \boldsymbol{U} + \partial_i \boldsymbol{F}^{i} = \boldsymbol{S}
\end{equation}

where the conserved variables \(\boldsymbol{U}\) are defined by

\begin{equation}
\frac{\boldsymbol{U}}{\sqrt{\gamma}}=\left(\begin{array}{c}
D\\
S_{j}\\
\tau
\end{array}\right)=\left(\begin{array}{c}
\rho W\\
\rho hW^{2}v_{j}\\
\rho hW^{2}-p-\rho W
\end{array}\right)
\end{equation}

and the fluxes as well as the sources are given by

\begin{equation}
\frac{\boldsymbol{F}^{i}}{\sqrt{\gamma}}=\left(\begin{array}{c}
D\left(\alpha v^{i}-\beta^{i}\right)\\
\alpha\tilde{S}_{\,j}^{i}-\beta S_{j}\\
\alpha\left(S^{i}-Dv^{i}\right)-\tau\beta^{i}
\end{array}\right)
\end{equation}

These conservation equations are implemented in the \texttt{Einstein
  Toolkit}'s \texttt{GRHydro} code \cite{Moesta13_GRHydro}, which uses high-resolution shock-capturing
methods.

We employ a fourth order Runge-Kutta method to advance in time,
either the WENO or PPM reconstruction method, second order flux
evaluations and either the HLLE or Marquina Riemann solver.  We set the Courant factor to 0.4.

Since we are not able to accurately track regions that transition from
the fluid to the vacuum regime, all simulations have an artificial
low-density background atmosphere which is evolved
freely. For all simulations presented in this study we choose an
atmosphere value of \(\rho_{\textrm{atm}} = 10^{-10}\text{M}^{-2}\approx 6.2\times 10^7 \textrm{g/cm}^{3}\) or \( \rho_{\textrm{atm}} = 10^{-14}  \, \text{M}^{-2}\approx 6.2\times 10^3 \textrm{g/cm}^{3}\).

\subsection{Initial Data}
\label{sec:initial-data}

We use a perturbed TOV solution as initial data. The matter perturbation
is implemented by adding the (2,2) pressure eigenfunction of an incompressible
Newtonian star in the Cowling approximation
\cite{Jones2002}. This form of the perturbation is
smooth and provides a well controlled framework to quantify numerical
accuracy.

\begin{figure}[h]
\centering
\includegraphics[width=\linewidth]{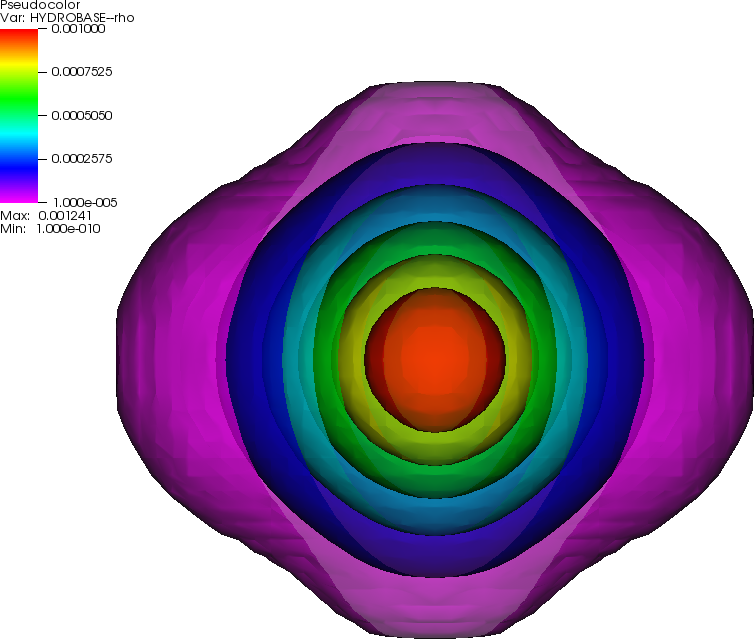}
\caption{Plot of density contours \(\rho\) (in code units) of a strongly
  perturbed TOV solution.  The star was excited using a pressure
  perturbation that was derived in the Cowling approximation to induce
  f-mode oscillations. The perturbation amplitude is chosen as
  \(\alpha=0.141\) for the star shown in the figure. }
\label{fig:perturbedtov}
\end{figure}

Specifically, the initial data was created with the \texttt{Einstein Toolkit}'s \texttt{TOVSolver} thorn.  Then, we added a pressure perturbation to excite this f-mode oscillations using the Cowling approximation.  The perturbation was given by
\begin{align}
\delta P=\alpha \rho \left(\frac{r}{R}\right)^2 Y_{22}
\end{align}
where \(\delta P\) is the perturbed pressure, \(\alpha\) is the amplitude of the perturbation, \(\rho\) is density, \(r\) is radial distance from the center of the star, \(R\) is the radius of the star, and \(Y_{22}\) is the \(l=2\) \(m = 2\) spherical harmonic. A plot of the density distribution resulting from this perturbation can be seen in Fig~\ref{fig:perturbedtov}.

\subsection{Equation of state}
\label{sec:EOS}

We adopt a polytropic equation of state 
\begin{align}
P&=K\rho ^\Gamma
\end{align}
where \(\Gamma\) is the adiabatic index and \(K\) is the constant of
proportionality.  For this work, we examined a polytrope with
\(\Gamma=2\) and \(K=100\) which is a common choice in the literature. We
leave to future work the expansion of the methods applied in this paper to other EOSs.

\subsection{Grid Setup}
\label{sec:grid}
The numerical grid is managed by the mesh refinement
\verb+Carpet+ \cite{Schnetter-etal-03b} driver for \verb+Cactus+
\cite{Goodale02a}. 

It implements a non-uniform grid via a nested set of movable boxes
(box-in-box) together with a hierarchical (Berger-Oliger-style)
timestepping. In these simulations we employ four refinement levels where
each additional level doubles the resolution of the enclosing one.  On
the finest grid, which is separately centered around each neutron star,
the resolution in each dimension is \(\Delta x = 0.0625 \text{M} \approx
93.75 \rm{m}\) for the highest resolution. 
 The physical domain extends to \(128 \text{M} \approx 192 \rm{km}\).

\subsection{Gravitational Waves}
\label{sec:grav-waves-theo}

The gravitational wave strain is extracted using the standard
Newman-Penrose formalism \cite{Newman62a}, in which a particular
contraction of the Weyl tensor with a suitably chosen null tetrad
yields a gauge-invariant quantity \(\psi_{4}\) that encodes the
outgoing gravitational radiation~\cite{Bishop2016}. Specifically, \(\psi_{4}\) is related to the
second time derivative of the two strain polarizations
\(\ddot{h}_{+,\times}\) by

\begin{equation}
\ddot{h}_{+} - {\rm i} \ddot{h}_{\times} = \psi_4
                                         = \sum_{\ell=2}^{\infty} \sum_{m=-\ell}^{\ell}
                                           \psi_4^{\ell m}\;_{-2}Y_{\ell m}(\theta,\varphi),
\end{equation}

where we also introduced the multipole expansion of \(\psi_{4}\) in spin-weighted
spherical harmonics \cite{Goldberg:1967} of spin-weight \(s=-2\).

\subsection{Models}
\label{sec:models}
We employed various models in which we altered simulation parameters such as perturbation amplitude, the spacetime formalism, etc.  We present and label these models in Table \ref{tab:Models}.  All models used a \(1.4\Msol\) TOV star which had a central density of \(0.00128\text{M}^{-2}\) and the EOS described in Section \ref{sec:EOS}.

\begin{table}[h]
  \caption{This table lists all of the models in the first column and the important quantities that changed between models.  These quantities were the spacetime formalism, the atmospheric density \(\rho_{\textrm{atm}}\), the perturbation amplitude \(\alpha\), the Riemann solver, and the reconstruction method. }
  \centering
  \resizebox{\linewidth}{!}{
    \begin{tabular}{llclll}

Model  & Formalism & \(\rho_{atm}\ (\text{M}^{-2})\) & \(\alpha\) & Riemann  & Recon.  \\
\midrule
PPM Big \(\alpha\)   & BSSN   & \(10^{-10}\) & 0.141 & Marquina  & PPM \\
PPM Big \(\rho_{\textrm{atm}}\) & BSSN      & \(10^{-10}\) & 0.0141   & Marquina  & PPM \\
PPM CCZ4   	& CCZ4 & \(10^{-10}\) & 0.0141    & Marquina  & PPM \\
PPM     	& BSSN & \(10^{-14}\) & 0.0141    & Marquina  & PPM \\
WENO    	& BSSN & \(10^{-14}\) & 0.0141    & Marquina  & WENO \\
WENO HLLE   & BSSN & \(10^{-14}\) & 0.0141    & HLLE		& WENO \\ 
\bottomrule
     \end{tabular}    }
  \label{tab:Models}\end{table}

\section{Results}
\label{sec:results}

\subsection{Extracted Gravitational Waveforms}
\label{sec:extracted_gravitational_waveforms}
\begin{figure}
\centering
\includegraphics[width=\linewidth]{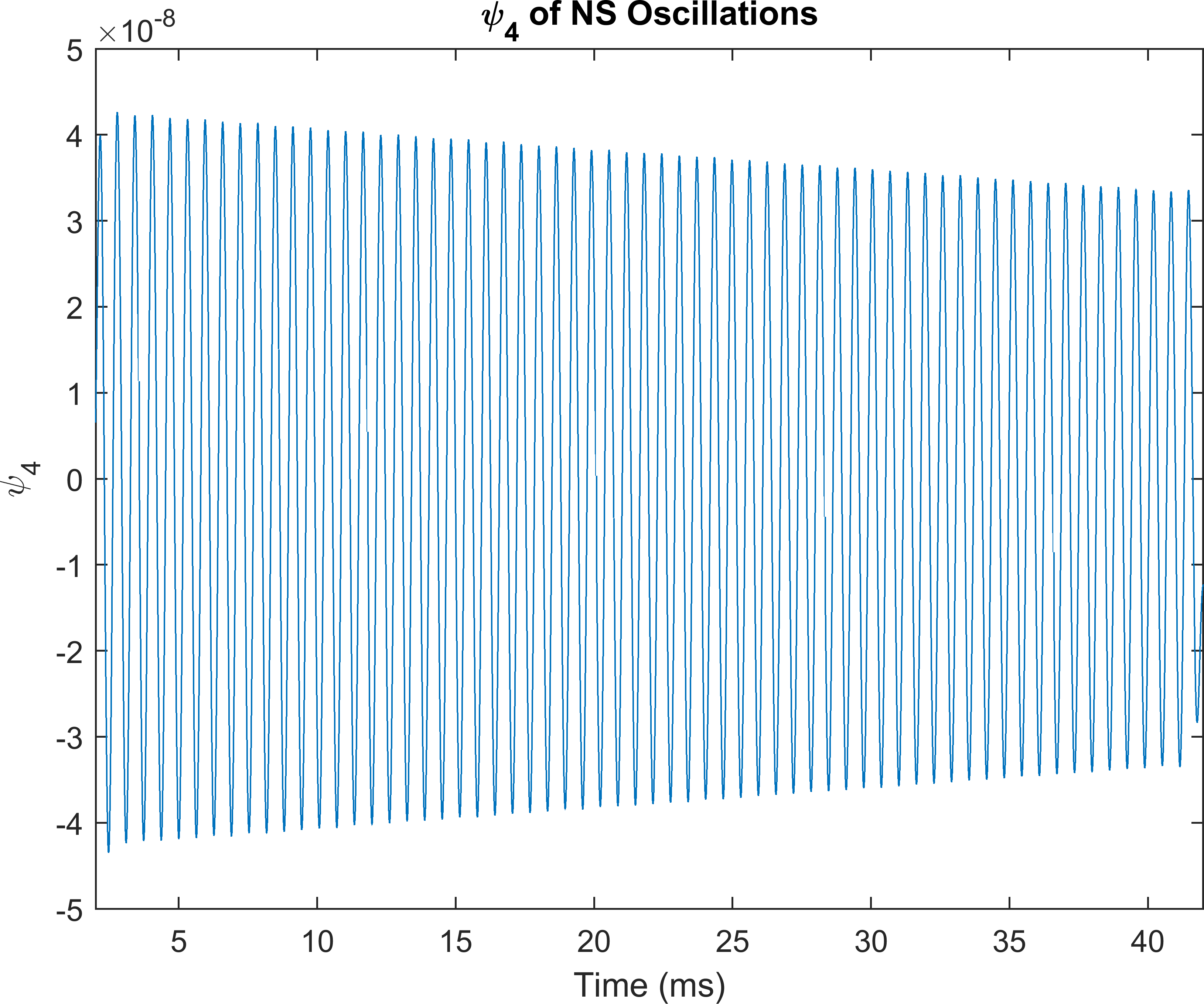}
\caption{The panel shows the real part of \(\psi_4\) from the PPM run with a resolution of 0.08\(\text{M}\). We observe that \(\psi_4\) takes the form of a
  decaying exponential. Furthermore, since the gravitational waves are 
  linearized polarized, we have confirmed that the imaginary part of \(\psi_4\) vanishes. }
\label{fig:psi4}
\end{figure}
We extracted the gravitational waves at distances of 
10\(\text{M} \), 40\(\text{M} \), 70\(\text{M} \), 100\(\text{M} \), and 120\(\text{M} \).  The analysis relied on the \(l=2\) \(m=2\) mode of the \( \psi_4 \) 
data extracted from 100\(\text{M}\) unless otherwise stated.

This system produced linearly polarized gravitational waves, which differs from other phenomena in gravitational wave physics which produce elliptically polarized waves.  This linear polarization arises from the lack of rotation in our oscillating neutron star.  In contrast, more often studied gravitational wave producing systems such as binary neutron star mergers and binary black hole mergers, use their rotation to produce their gravitational waves and give their gravitational waves their elliptical polarization. 

The polarization is imprinted in the real and imaginary parts of \( \psi_{4} \) 
which represent the plus and cross polarizations respectively.  The real part exhibits a sinusoidal form with a frequency close to the fundamental oscillation mode frequency and on top a secular exponential decay, see Fig~\ref{fig:psi4}. The imaginary part is zero to almost machine precision, indicating that no cross polarization was present.

We note that \(\psi_{4}\) had
high frequency noise as a result from passing through refinement
layers. We removed these high frequency features with low pass filters that are described in detail in Appendix~\ref{sec:filtering}.

\subsection{Convergence}
\label{sec:convergence}
\begin{figure}
\centering
\includegraphics[width=\linewidth]{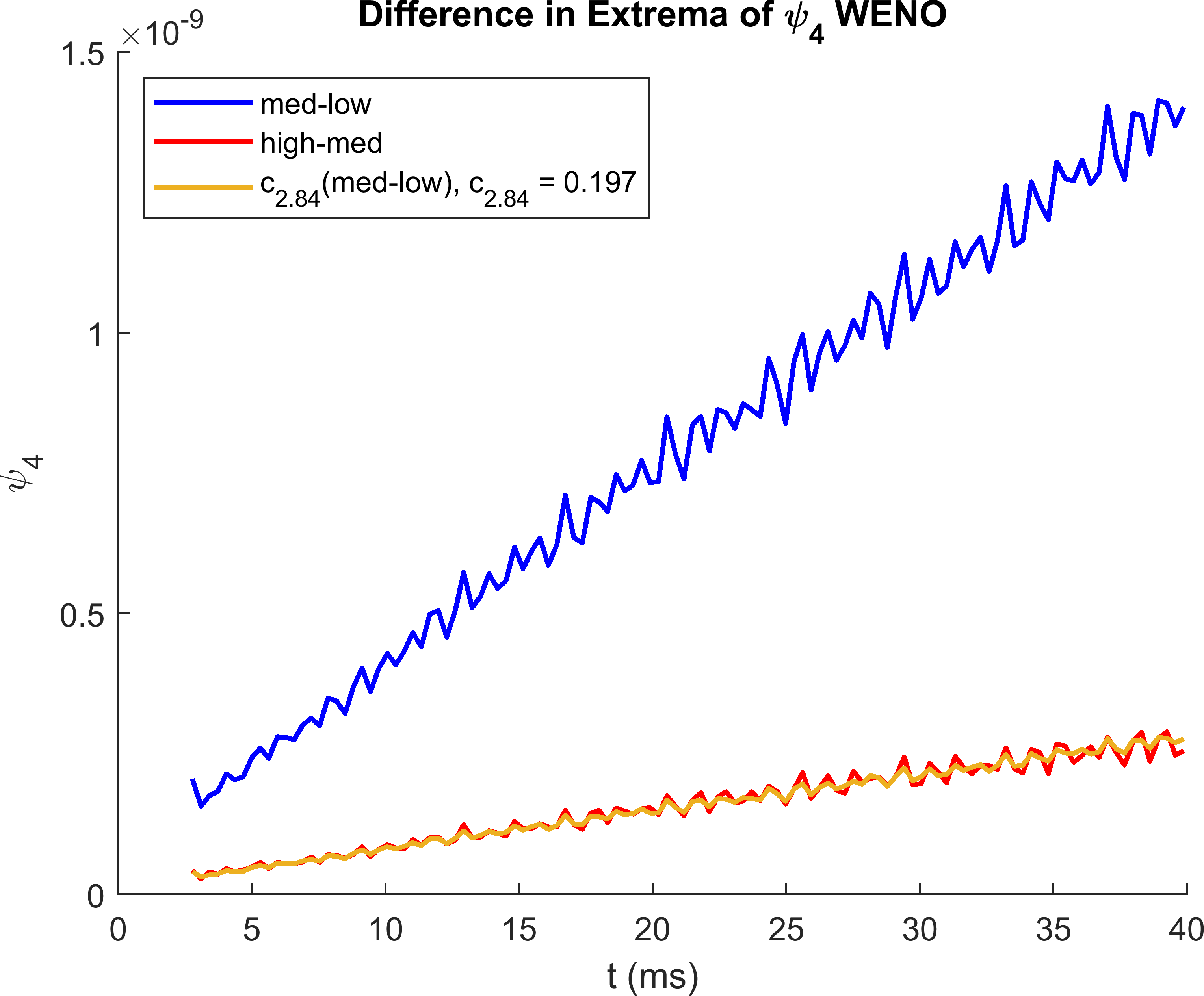}
\caption{Convergence of the extrema of
  \(\psi_4\). Here low, med, high refer to runs with fine resolutions of \(0.125\text{M}\), \(0.08\text{M}\) and, \(0.0625\text{M}\) respectively for the WENO case. This information can be inferred by noticing that 
  the resolutions satisfy the relation \((\textrm{high}-\textrm{med}) = c_n(\textrm{med}-\textrm{low})\), see Eq.~\eqref{eq:conv}.
  We found \(c_n \sim 0.197\) by taking \(c_n\) to be the best fit of our data to Eq.\eqref{eq:conv}. Using Eq.~\eqref{eq:c_fac} for \(c_{2.84} \sim 0.197 \), we found that \(n=2.84\)  which is consistent with the assumed order of convergence. This result is close to the order of our lowest
  order contributions to our numerical method, which demonstrates both the convergence and internal 
  consistency of our numerical methods. We acknowledge that the value in this plot
  differs from that quoted in Table \ref{tab:Results} as this convergence test
  applies to a longer portion of the waveform.  The value quoted in Table
  \ref{tab:Results} only uses the portion of the waveform that was used to compute
  the damping times.  See Section \ref{sec:damping_evol} for more details.}
\label{fig:conv}
\end{figure}

We demonstrate that differences between resolutions behave as expected
given the numerical method we adopt. This implies that the numerical
error in our simulations can be quantified. Only with such a reliable
error estimate is it possible to tell whether any damping in the
gravitational waves is physically caused by gravitational radiation or by
numerical dissipation. There are several possibilities to demonstrate
convergence in the GW output of such models. We choose to perform
convergence tests on the extrema of \(\psi_{4}\).

As usual, we assume that the numerical error as represented by a Taylor
expansion is dominated by the leading order term. In this case three
resolutions suffice to demonstrate convergence and fit the parameters of
the Taylor expansion. Specifically, the numerical solution for high,
medium, and low resolutions are related by
\begin{align}
\label{eq:conv}
\psi_{4,\textrm{high}}-\psi_{4,\textrm{med}}=c_n(\psi_{4,\textrm{med}}-\psi_{4,\textrm{low}})\,,
\end{align}
\noindent where \(c_n\) is the convergence factor of order \(n\) and is determined
only by the resolutions and convergence order as
\begin{align}
\label{eq:c_fac}
c_n=\frac{\Delta x_{\textrm{high}}^n - \Delta x_{\textrm{med}}^n}{ \Delta x_{\textrm{med}}^n -
  \Delta x_{\textrm{low}}^n}.
\end{align}

\begin{figure}
\centering
\includegraphics[width=\linewidth]{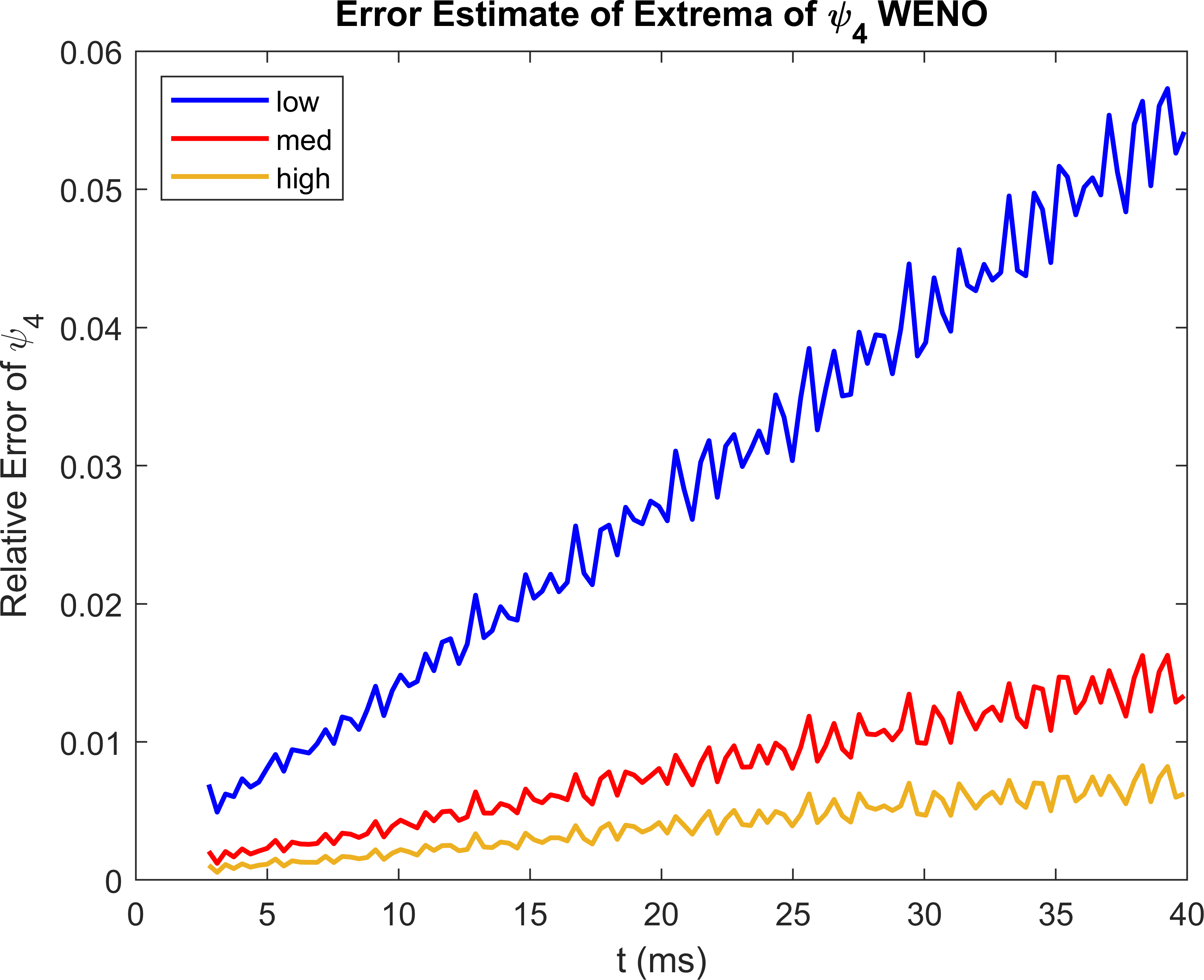}
\caption{Estimate for the relative error in \(\psi_4\). Here low, med, high refer to runs with fine resolutions of \(0.125\text{M}\), \(0.08\text{M}\) and, \(0.0625\text{M}\) respectively for the WENO case.
This error estimate was obtained using Eq.~\eqref{eq:error} and dividing by the value of \( \psi_{4} \) at each instant in time. Unlike Fig~\ref{fig:conv} where we used a constant value of \(c_n\) for a best fit to Eq.~\eqref{eq:conv},
our value of \(c_n\) and \(n\) here are solutions to Eqs.~\eqref{eq:conv} and ~\eqref{eq:c_fac} at each point in time.  }
\label{fig:error}
\end{figure}

\noindent We present the convergence order for all our numerical relativity simulations in Table~\ref{tab:Results}. 
A sample case of this analysis for the WENO reconstruction is presented in Fig~\ref{fig:conv}. Using the convergence factor and order, we were able
to estimate the systematic error due to the changing resolution as
\begin{align}
\textrm{error}=|c_n \Delta x^n|\,.
\label{eq:error}
\end{align}
\noindent The results for the estimated error of the WENO reconstruction is presented in Fig~\ref{fig:error}.  This error behaves as systematic error caused by the differences in simulation resolution.  As such, we did not propagate this error into the analysis because the analysis aims at characterizing the effect of this same resolution dependence on the damping times.

\subsection{Fourier Analysis}
\label{sec:fft}
\begin{figure}
\centering
\includegraphics[width=\linewidth]{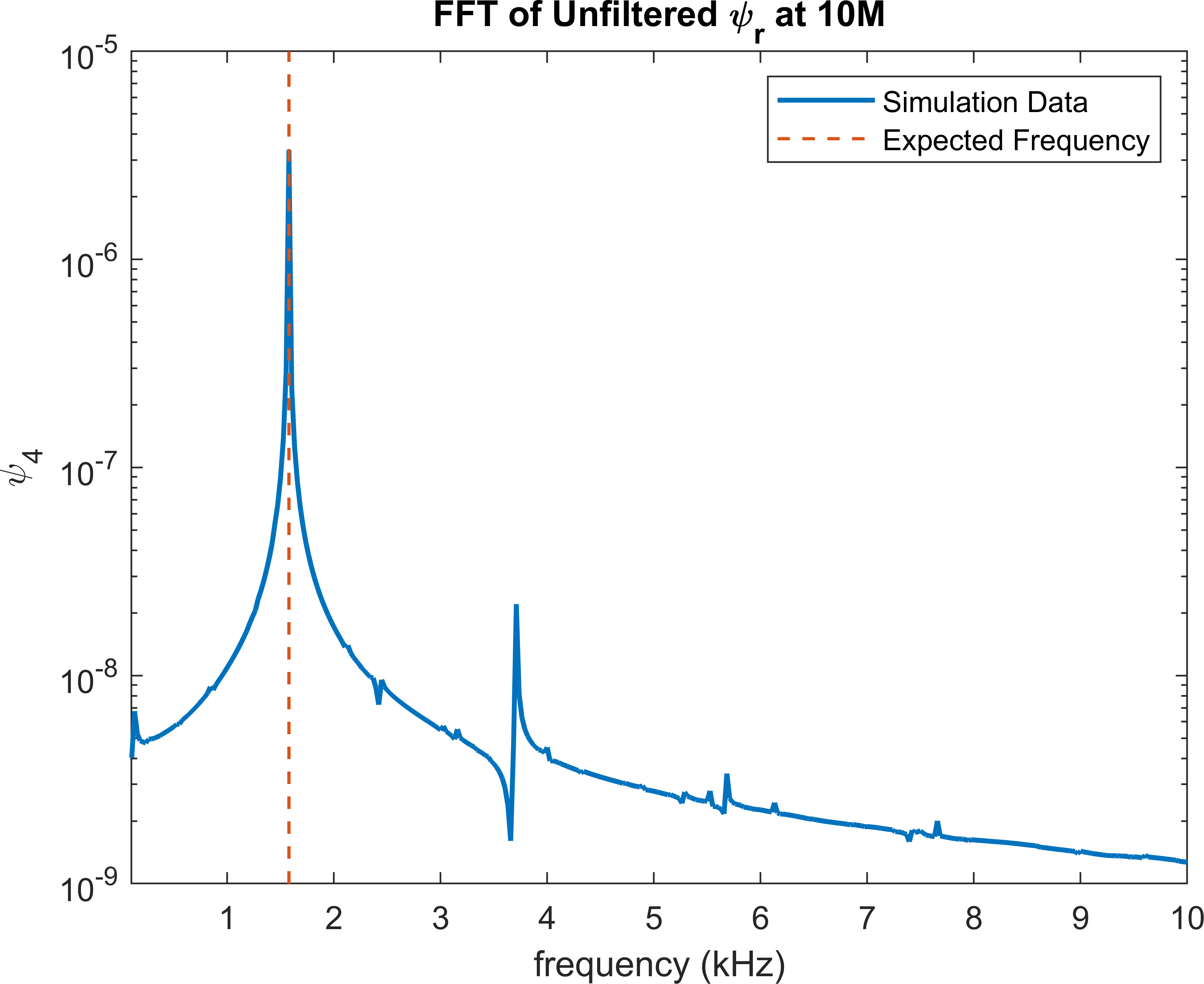}
\caption{Fast Fourier Transform (FFT) of the \(\psi_4\) data from WENO HLLE simulation with a fine resolution of 0.0625\( \text{M} \).
  The gravitational waves here were extracted from 10M.  The 10M data best illuminates the excitations in the frequency spectrum because it occurs before any AMR boundary introduces high frequency features (we also present the frequency spectrum extracted at 100M in Figure~\ref{fig:filtering_100M}). We note the primary is near the expected f-mode excitation frequency of 1.579 kHz.  We also observe secondary excitations at 3.710 kHz, 5.684 kHz, and 7.658 kHz which are consistent with the first overtones (p-modes) of the f-mode.  We note that the spikes at the higher frequencies are suppressed after applying the filtering. See Appendix~\ref{sec:filtering} for more details.}
\label{fig:fft}
\end{figure}

Frequency spectra of the gravitational waveform were examined by taking
the Fast Fourier Transform (FFT) of the \(\psi_{4}\) data. We plot the frequency spectrum of one of the runs in Fig~\ref{fig:fft}.  We see that
the most excited frequency
is nearly aligned with the expected frequency of 1.579 kHz, illustrating that our perturbation excited the desired f-mode.  In addition, we observe at least three secondary excitations which we believe to be p-modes.

\subsection{Frequency}
\label{sec:frequency}
We also investigated the frequency {\it evolution} as the run progressed.
Therefore, we took the roots of $\psi_{4}$ time series to estimate
the period, which was in turn used to obtain frequency estimates for each
cycle.  We used the average frequency as the value for the simulation and
the standard deviation as the error.

\indent
Although most of the runs exhibited constant frequency regardless of
resolution, runs with high density atmospheres \(10^{-10}\text{M}^{-2}\) of and a
length longer than 10\ms{} experienced a rise in frequency as runs
progressed.

\begin{figure}
\centering
\includegraphics[width=\linewidth]{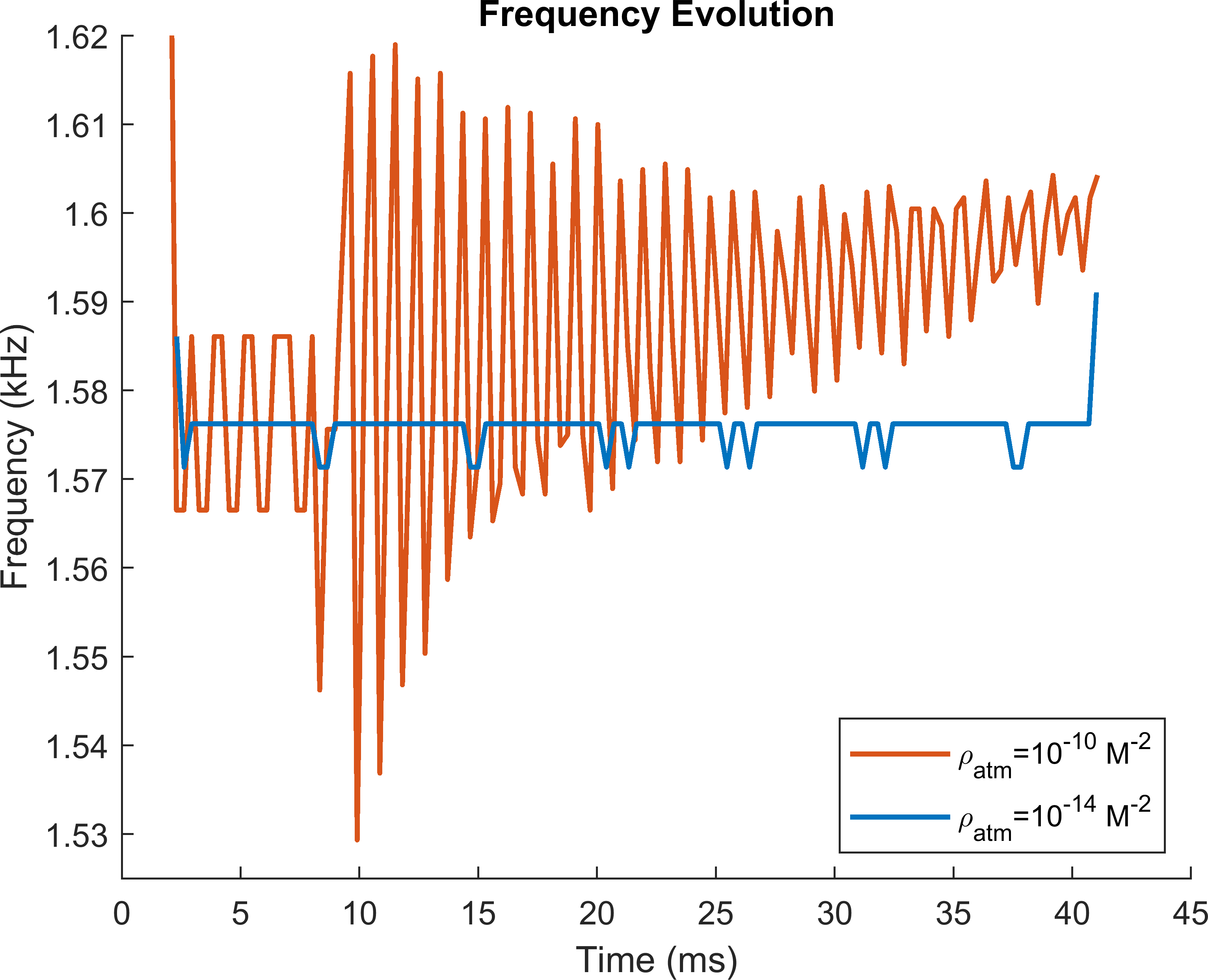}
\caption{Frequency evolution of the waveform as time progressed of the PPM and PPM Big $\rho_{\textrm{atm}}$ runs at a fine resolution of 0.125$\text{M}$.  This frequency was computed by examining the roots of the waveform time series and finding the time interval between them.  This allows us to track the frequency evolution throughout the run.  For the high atmosphere plot in red, the frequency increases after about 10\ms{}.  The low atmosphere run in blue remained constant throughout the entirety of the run. }
\label{fig:freq_evolution}
\end{figure}

In contrast, the frequency of low density atmosphere runs
of \(10^{-14}\text{M}^{-2}\) remained constant for even the longest runs
which laster 50\ms{}.  We can see the frequency evolution in
Fig~\ref{fig:freq_evolution}.  The likely physical cause of this behavior
is accretion of artificial atmosphere material onto the star which leads
to a steady change in the background stellar model and hence a slight
change in its fundamental oscillation frequencies.

\indent 

We found that the frequency was fairly independent of resolution as
displayed in Table \ref{tab:Results}.  Though the higher amplitude runs
did experience a lower frequency than the lower perturbation runs. The
frequencies all appear to be around 1.579 kHz which is the expected
frequency for f-mode oscillations based on linear theory.

\indent
We note that since the gravitational waves were linearly polarized, we could not determine instantaneous frequency by
taking the derivative of the phase of the gravitational waves as the
imaginary part was nearly zero.

\subsection{Damping Times}
\label{sec:damping_times}

\begin{figure}
\centering
\includegraphics[width=\linewidth]{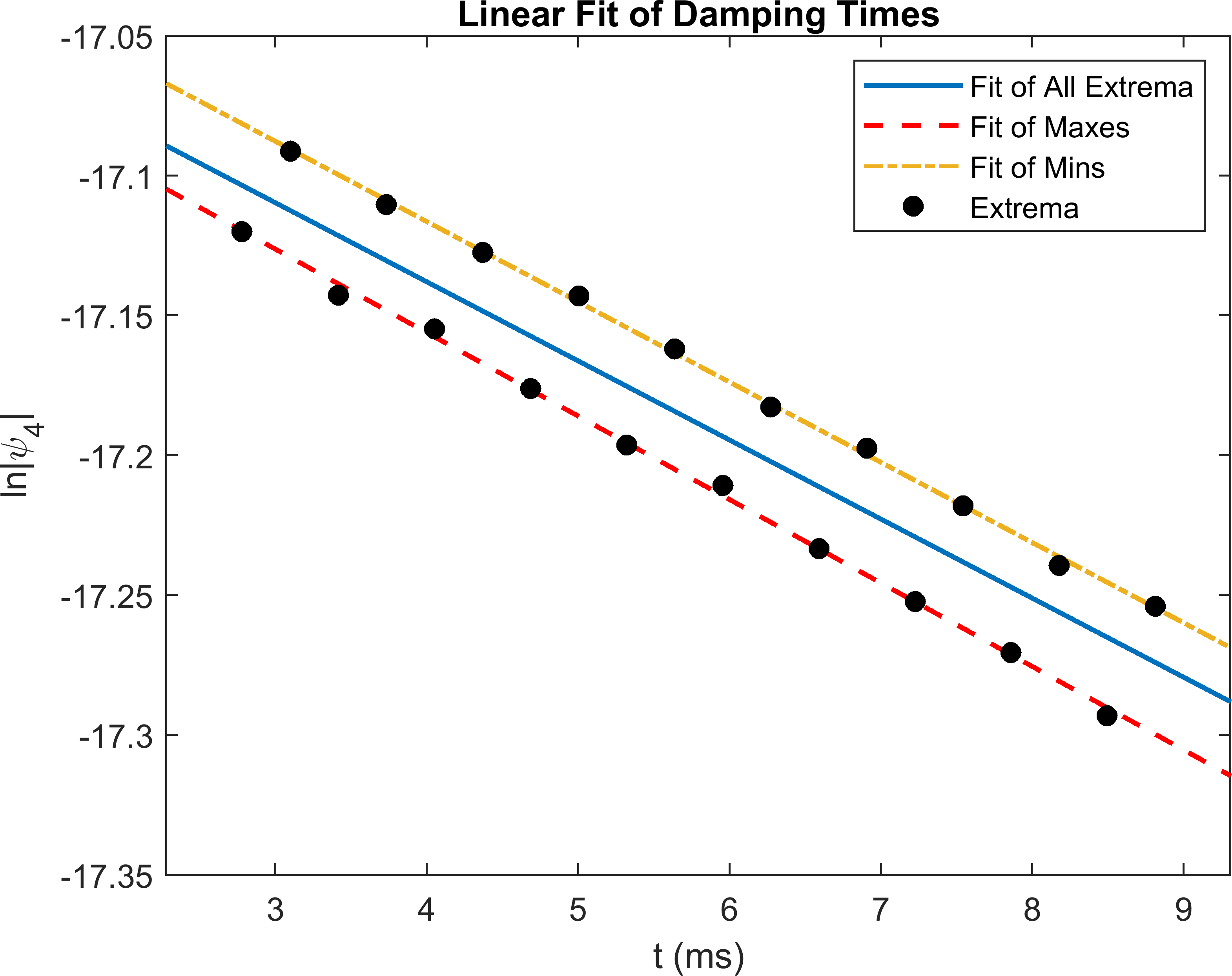}
\caption{Linear fit of the extrema of \(\psi_4\) that was used to obtain
  damping times of the gravitational waves for the PPM CCZ4 simulation with a fine resolution of 0.16$\text{M}$. We note in particular that
  the extrema of $\psi_4$ differed by an offset that may skew the
  results. Therefore, we examined the slopes computed using the maxima
  and minima individually to provide an error estimate for the damping
  time of each simulation. The quoted value for the damping times were
  computed using the full dataset of extrema.}
\label{fig:lin_fit}
\end{figure}

\begin{figure*}
\centering
\includegraphics[width=\linewidth]{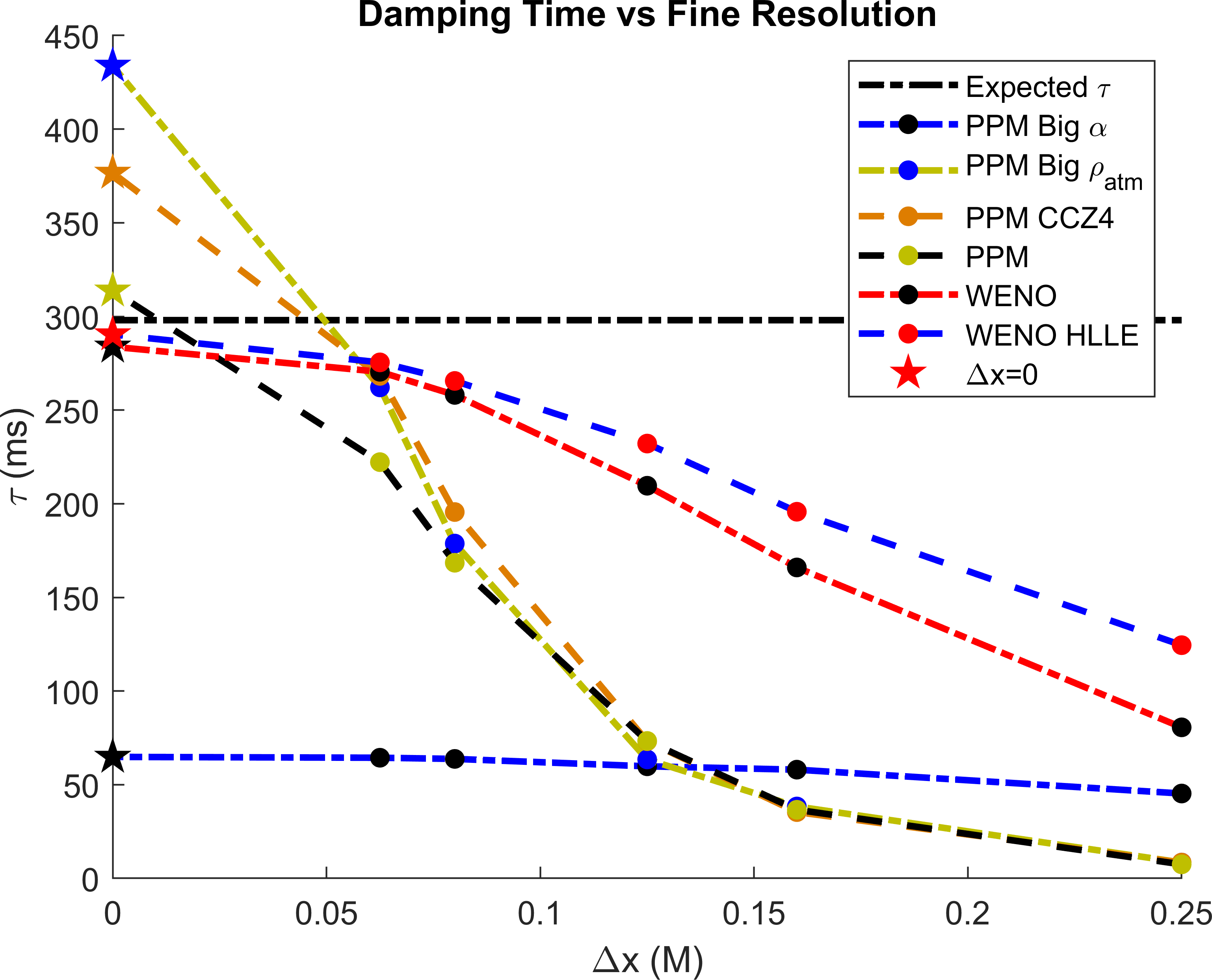}
\caption{
Damping times for simulations of neutron star f-mode
oscillations as a function of fine resolution. The black dash-dot line represent
the expected damping time for a static TOV of this mass computed using
linear theory.  The star value at \(\Delta x\rightarrow 0 \text{M}\) is the value
obtain for the frequencies by extrapolating to infinite resolution.  We
notice that the runs using PPM reconstruction behaved similarly to one
another.  However, the high atmosphere PPM \(
\rho_{\textrm{atm}} \) and PPM CCZ4 runs diverge from the PPM run at finer resolutions.  This causes the Richardson extrapolated value of those high
atmosphere sequences to significantly overshoot the expected damping time from
linear theory.  We described in Section \ref{sec:frequency} that the high
atmosphere runs experience increased frequency as the run progressed. We believe 
the effects of this evolving frequency propagate into the damping time
calculation, leading to the divergence of these sequences of damping times.
The WENO reconstruction runs also exhibited similar characteristics to each
other. In addition, the WENO runs are significantly more accurate than the 
PPM runs at a given resolution. PPM models required twice the resolution sed in the WENO models for similar accuracy. The high perturbation run
exhibited entirely different damping time characteristics from the others, indicating that it was likely not in the linear perturbation regime.  }
\label{fig:damping_times}
\end{figure*}

Damping times were extracted by performing linear fits to the natural
log of the extrema of \(\psi_{4}\), as shown in Fig~\ref{fig:lin_fit}. Error estimates were computed using the 
standard deviations of the damping time calculations using the extrema 
of \(\psi_{4}\) individually rather than from all the extrema in the simulation dataset. 

To compute the damping times from our suite of numerical relativity simulations we used 
the WENO and PPM reconstructions methods, finding that the former is much more robust 
and accurate than the latter. Since simulations with \(\Delta x<0.0625\text{M}\) were deemed to be too computationally 
intensive to run, we extrapolated these results to the continuum limit to see if they approach the expected value of 298\ms{}.  We used Richardson extrapolation with unequal resolution ratios to estimate the value at infinitely fine resolution. A summary of these calculations is presented in Table \ref{tab:Results}. 

Fig~\ref{fig:damping_times} shows that the WENO reconstruction approaches the 
linear perturbation prediction of 298\ms{} for the highest resolution run of each simulation. 
The three simulations that significantly depart from the linear perturbation
regime correspond to two categories: (i) the two simulations with damping times much larger than 298\ms{} 
correspond to simulations with large artificial atmospheres, \(\rho_{\textrm{atm}}\sim 10^{-10}\text{M}^{-2}\); 
(ii) the simulation with estimated damping time much smaller than 298\ms{} is clearly outside of the 
linear perturbation regime by construction. 

These results indicate that provided a numerical setup in which the atmosphere used for the 
simulations satisfies \(\rho_{\textrm{atm}}\lesssim 10^{-14}\text{M}^{-2}\), and the perturbation applied 
to the TOV star is within the linear regime, the extracted damping times with the WENO reconstruction method from our suite of 
numerical simulations is consistent with linear perturbation predictions.

\begin{table}
\caption{Table depicting the results our from oscillating neutron star simulations.  These results include the resolution of the simulation \(\Delta x\), its frequency \(f\), the error in that frequency \(\Delta f\), the damping time \(\tau\) and its estimated error \(\Delta \tau\).  The damping error is the statistical fitting error of the data and does not include any systematic errors alluded to in Sections \ref{sec:convergence} and \ref{sec:damping_evol}.
 The table is divided by the simulation models described in Table \ref{tab:Models}. Some cases yield convergence orders that exceed the formal accuracy of the numerical schemes involved. Such so-called overconvergence usually indicates that the data is close to but not quite within the convergent regime. At the end of each group of simulations, we include the damping time value at \(\Delta x\rightarrow0\) using Richardson extrapolation. This procedure will systematically underestimate the error in presence of over-convergence and should be taken with a grain of salt.  We did not include errors on this value nor the frequency when performing this extrapolation to infinitely fine resolution.  The expected frequency and damping time for all simulations in the table according to linear theory are 1.579 kHz and 298\ms{} respectively.
}

\resizebox{\linewidth}{!}{
\begin{tabular}{lrccrcc}
 & \(\Delta x\) (\(\text{M}\)) & \(f\) (kHz) & \(\Delta f\) (kHz) & \(\tau\) (ms) & \(\Delta\tau\) (ms) & \(n\) \\
\midrule
PPM Big \(\alpha\) & 0.25 & 1.551 & 0.023 & 45.29 & 0.39 & 4.84 \\
 & 0.16 & 1.552 & 0.013 & 58.02 & 0.93 & 4.84 \\
 & 0.125 & 1.555 & 0.017 & 59.81 & 1.65 & 4.84 \\
 & 0.08 & 1.555 & 0.013 & 63.76 & 1.42 & 4.84 \\
 & 0.0625 & 1.555 & 0.011 & 64.39 & 1.46 & 4.84 \\
 & 0 & - & - & 64.83 & - & 4.84 \\
PPM Big \(\rho_{\textrm{atm}}\) & 0.25 & 1.572 & 0.019 & 8.44 & 0.21 & 3.83 \\
 & 0.16 & 1.579 & 0.014 & 38.36 & 1.56 & 3.83 \\
 & 0.125 & 1.580 & 0.023 & 63.48 & 9.88 & 3.83 \\
 & 0.08 & 1.580 & 0.004 & 178.73 & 3.79 & 3.83 \\
 & 0.0625 & 1.581 & 0.004 & 262.11 & 7.58 & 3.83 \\
 & 0 & - & - & 433.80 & - & 3.83 \\
PPM CCZ4 & 0.25 & 1.574 & 0.004 & 8.41 & 0.13 & 3.14 \\
 & 0.16 & 1.577 & 0.003 & 35.36 & 0.96 & 3.14 \\
 & 0.125 & 1.579 & 0.004 & 73.21 & 0.57 & 3.14 \\
 & 0.08 & 1.580 & 0.004 & 195.57 & 3.76 & 3.14 \\
 & 0.0625 & 1.581 & 0.004 & 268.25 & 6.33 & 3.14 \\
 & 0 & - & - & 376.65 & - & 3.14 \\
PPM & 0.25 & 1.573 & 0.005 & 7.47 & 0.16 & 2.98 \\
 & 0.16 & 1.576 & 0.004 & 36.61 & 0.55 & 2.98 \\
 & 0.125 & 1.576 & 0.003 & 73.35 & 1.05 & 2.98 \\
 & 0.08 & 1.577 & 0.004 & 168.48 & 2.71 & 2.98 \\
 & 0.0625 & 1.577 & 0.003 & 222.20 & 3.85 & 2.98 \\
 & 0 & - & - & 313.74 & - & 2.98 \\
WENO & 0.25 & 1.573 & 0.005 & 80.61 & 0.99 & 2.79 \\
 & 0.16 & 1.576 & 0.004 & 165.98 & 1.54 & 2.79 \\
 & 0.125 & 1.577 & 0.004 & 209.60 & 2.14 & 2.79 \\
 & 0.08 & 1.577 & 0.003 & 258.09 & 4.07 & 2.79 \\
 & 0.0625 & 1.578 & 0.003 & 270.44 & 4.80 & 2.79 \\
 & 0 & - & - & 283.80 & - & 2.79 \\
WENO HLLE & 0.25 & 1.574 & 0.005 & 124.48 & 8.42 & 5.77 \\
 & 0.16 & 1.576 & 0.004 & 195.72 & 1.71 & 5.77 \\
 & 0.125 & 1.577 & 0.003 & 232.12 & 2.44 & 5.77 \\
 & 0.08 & 1.579 & 0.011 & 265.56 & 2.80 & 5.77 \\
 & 0.0625 & 1.578 & 0.003 & 275.47 & 5.00 & 5.77 \\
 & 0 & - & - & 290.37 & - & 5.77 \\
\bottomrule
 \end{tabular}
}
\label{tab:Results}
\end{table}

\subsection{Evolution of Damping Time}
\label{sec:damping_evol}
\begin{figure}
	\centering
	\includegraphics[width=\linewidth]{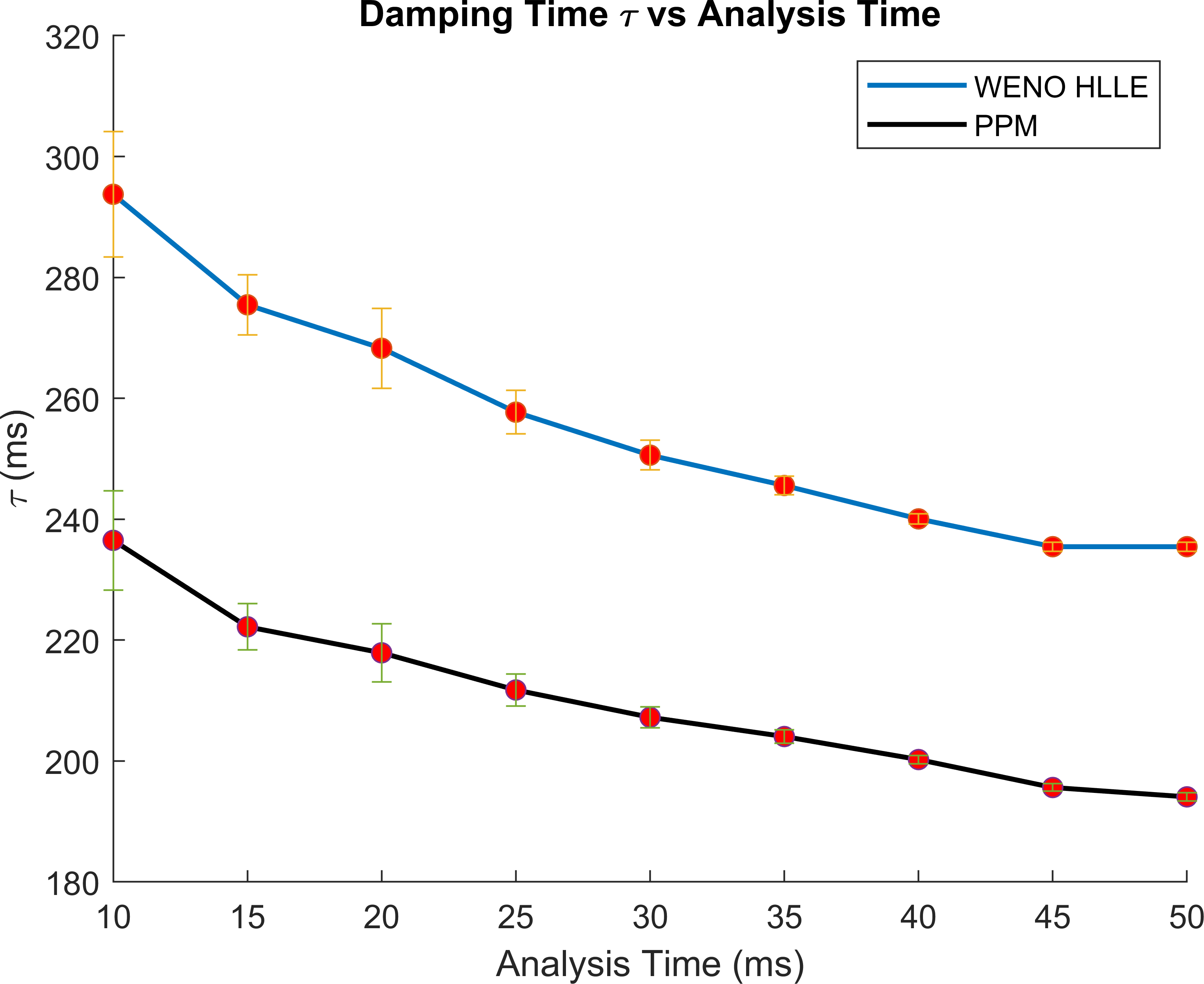}
	\caption{Plot of the damping time $\tau$ vs analysis time with the statistical fitting error at each data point for the WENO HLLE and PPM data sets with $\Delta x=0.0625\text{M}$.  The analysis time is defined as the cutoff time for waveform data to be used in the analysis.  Data beyond the analysis time was discarded for the purpose of the analysis.  The first 2 ms of data which is included in the analysis was also discarded to allow for any transient effects to decay. We notice that statistical error clearly decreases as the analysis time increases.  However, we observe that $\tau$ decreases away from the expected value of 298 ms with increased analysis time.  We noticed that in both curves in the plot that there appears to be a local minimum in the statistical uncertainty at an analysis time of 15.  As such we selected, an analysis time of 15 ms for the damping time analysis in this paper. }
	\label{fig:tau_vs_time}
\end{figure}
High resolution runs with longer damping times required longer analysis times to obtain satisfactory statistical fits. This necessity derives from the difference in position of individual points being significant compared to the difference induced by the true damping of the waveform over short timescales.  In essence, the longer we run the simulation, the closer we get to running for the damping time we would like to measure; thereby, we reduce the statistical fitting error in the damping time.\\
\indent
Although increasing the run time decreases the statistical error in the damping time, it introduces systematic error in the damping time calculation.  We found that as the run progressed, the waveform did not exhibit a true exponential decay.  Instead, the waveform systematically decayed quicker than an exponential damping time would suggest at long times.  This posits the existence some other source of decay in the waveforms, most likely some form of numerical dissipation.\\
\indent
Fig~\ref{fig:tau_vs_time} illustrates the evolution of the damping time value vs the analysis time, the cutoff time for waveform data to in the analysis to calculate the damping time beyond which the data would not be used for such analysis.  We also note that the first 2 ms of data also was not used in the analysis to allow time for any transients to die out. In addition, the figure depicts the statistical fitting error at each of those points.  One can observe that the damping time and statistical error decrease as the analysis time increase.  We observed a local minimum in the statistical error at an analysis time of 15 ms and thus chose that analysis time for the computation of the damping times in Table \ref{tab:Results}.

\section{Conclusions}
\label{sec:conclusions}

We have carried out systematic analyses of the accuracy needed to extract
the frequency and damping time of f-mode oscillations. We have
demonstrated the robustness of our
results by extracting these observables, and demonstrating that our 
results are similar when using both the WENO and PPM reconstruction methods.

We have shown that for small perturbations, the damping times extracted from 
our numerical relativity simulations are 
consisted with linear theory perturbations when we consider atmospheres 
\(\rho_{\textrm{atm}}\leq10^{-14}\text{M}^{-2}\). On the other hand, we have found 
that if artificial atmospheres are poorly chosen, \(\rho_{\textrm{atm}}\leq10^{-10}\text{M}^{-2}\), 
the extracted damping times from our numerical simulations differ significantly from 
linear perturbation predictions at evolutions \( \gtrsim 10 \)ms. 
This may be caused due to accretion of the 
atmosphere, which introduces a secular change (increased mass) in the background solution.

The convergence analyses we present suggest that for a careful choice 
of numerical schemes a minimum resolution of 
\(\sim 0.125\text{M}\) is needed to accurately extract the GW induced damping time 
of f-mode oscillations. WENO reconstruction clearly outperforms the PPM
scheme, which needed twice the resolution to obtain similar accuracies in
the damping times. This resolution is within the range of existing neutron star
numerical relativity simulations that have been presented in the
literature albeit on the more computationally expensive end.

Having established a blueprint of numerical accuracy to extract the frequency and damping times of 
f-modes, in future work we will put this formalism to use and investigate the accuracy with which 
we can infer properties of the EOS through GW observations of highly eccentric neutron star encounters.

\section*{Acknowledgements}
This research is part of the Blue Waters sustained-petascale computing project, which is supported by the National Science Foundation (awards OCI-0725070 and ACI-1238993) and the State of Illinois. Blue Waters is a joint effort of the University of Illinois at Urbana-Champaign and its National Center for Supercomputing Applications. Some numerical results were obtained using the University of Maryland Deepthought2 HPC Cluster. This work was supported in part by joint research workshop award 2015/50421-8 from FAPESP and the University of Maryland, College Park. This work was supported in part by Perimeter Institute for Theoretical Physics. Research at Perimeter Institute is supported by the Government of Canada through the Department of Innovation, Science and Economic Development Canada and by the Province of Ontario through the Ministry of Research, Innovation and Science. We acknowledge support from the NCSA. NSF-1550514 and NSF-1659702 grants are gratefully acknowledged. We thank the \href{http://gravity.ncsa.illinois.edu}{NCSA Gravity Group} for useful feedback.

\bibliography{references,aeireferences,refs}
\bibliographystyle{apsrev4-1}

\clearpage
\appendix

\section{Filtering}
\label{sec:filtering}
\begin{figure*}
\centerline{
	\includegraphics[width=0.5\linewidth]{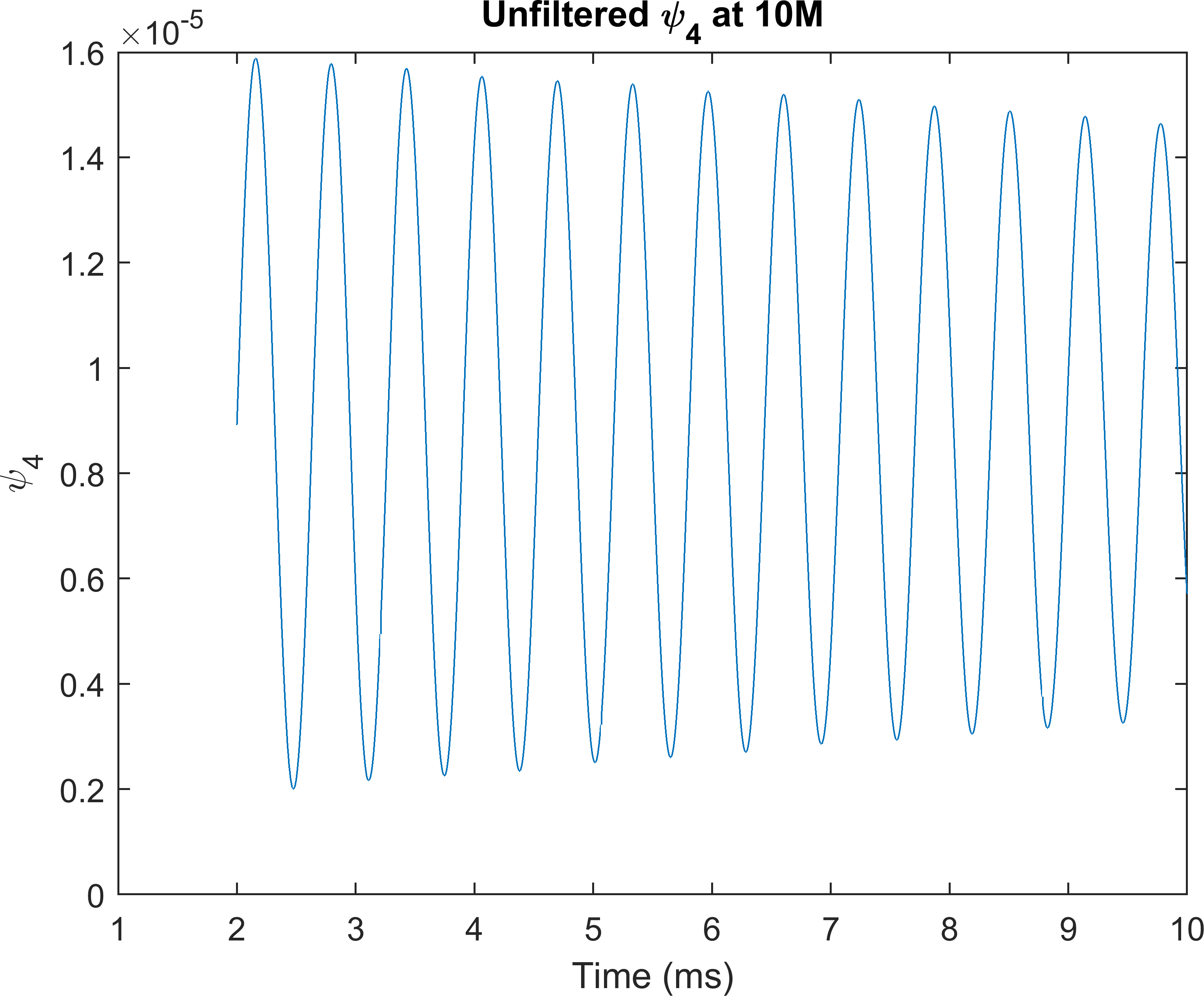}
	\includegraphics[width=0.5\linewidth]{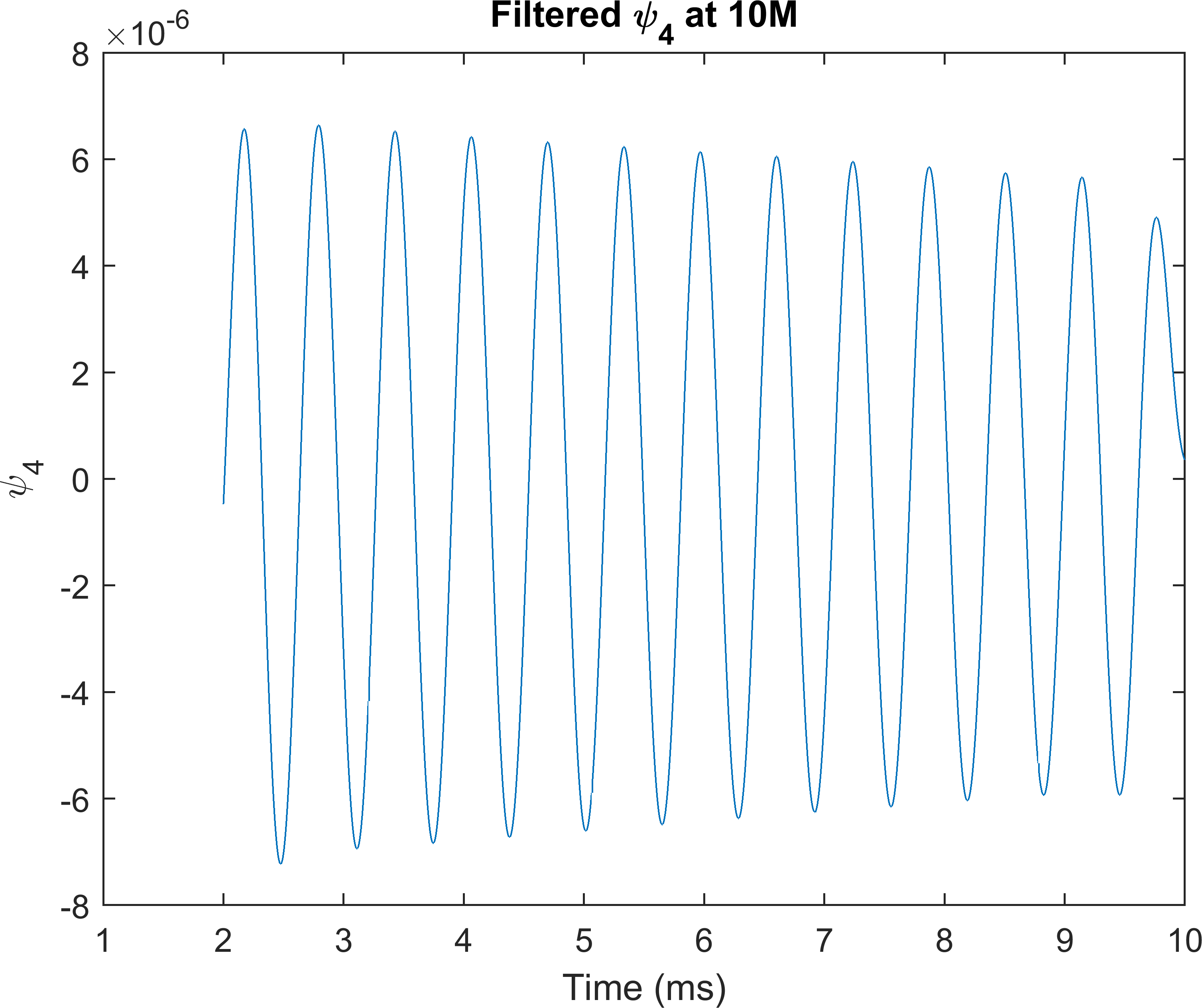}
	}
	\caption{Plot of the $\psi_4$ data extracted at 10$\text{M}$. The left plot depicts the unfiltered data. We note that the unfiltered data lacks any sort high frequency component as it has yet to pass through a refinement boundary.  When we filter the data as in the right plot, we observe that the waveform is now centered at the origin rather than offset as it had been previously.  The waveform is also hardly distorted, with most of the distortion occurring at the end of the waveform. }
	\label{fig:filtering_10M}
\end{figure*}

\begin{figure*}
\centerline{
	\includegraphics[width=0.5\linewidth]{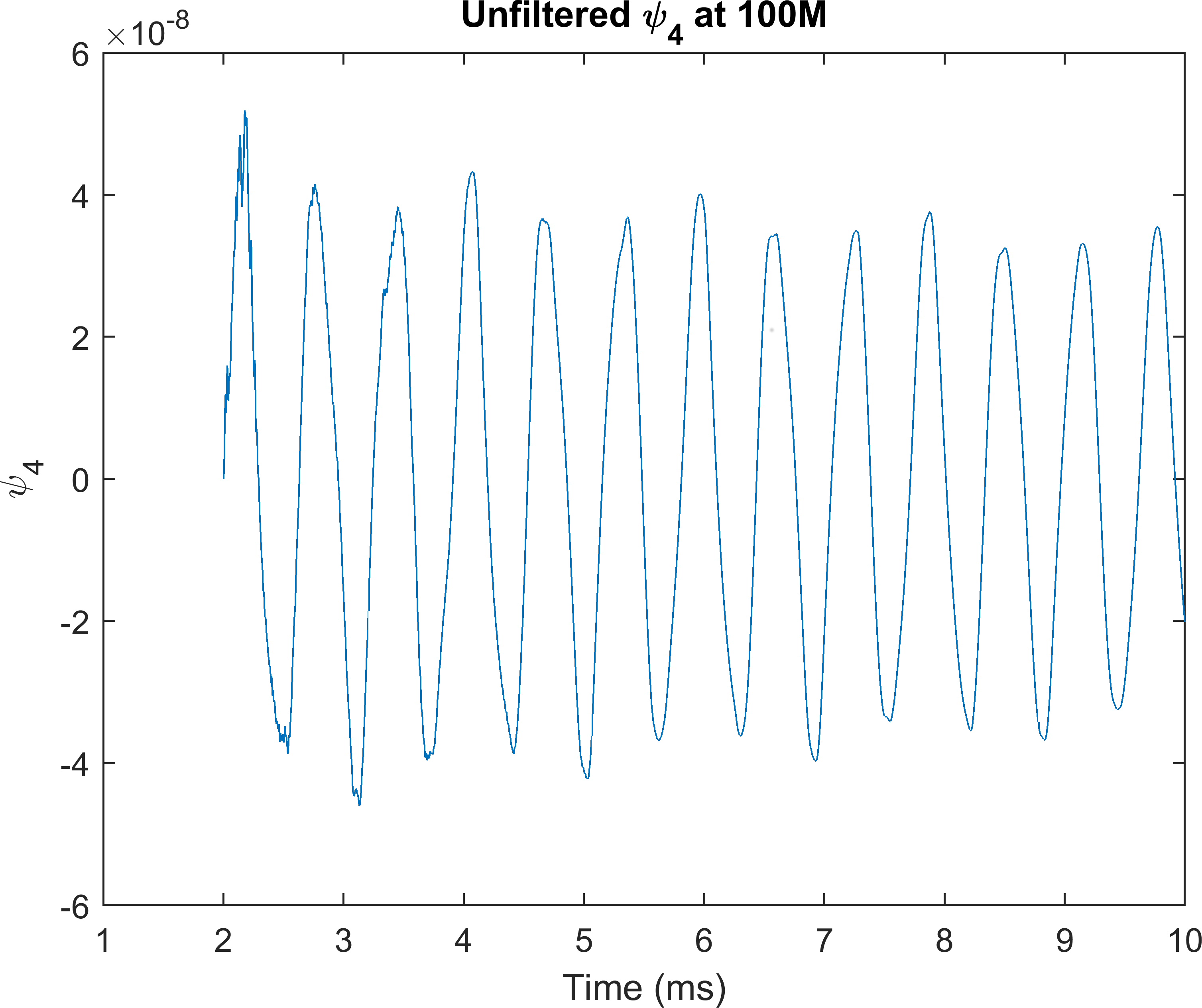}
	\includegraphics[width=0.5\linewidth]{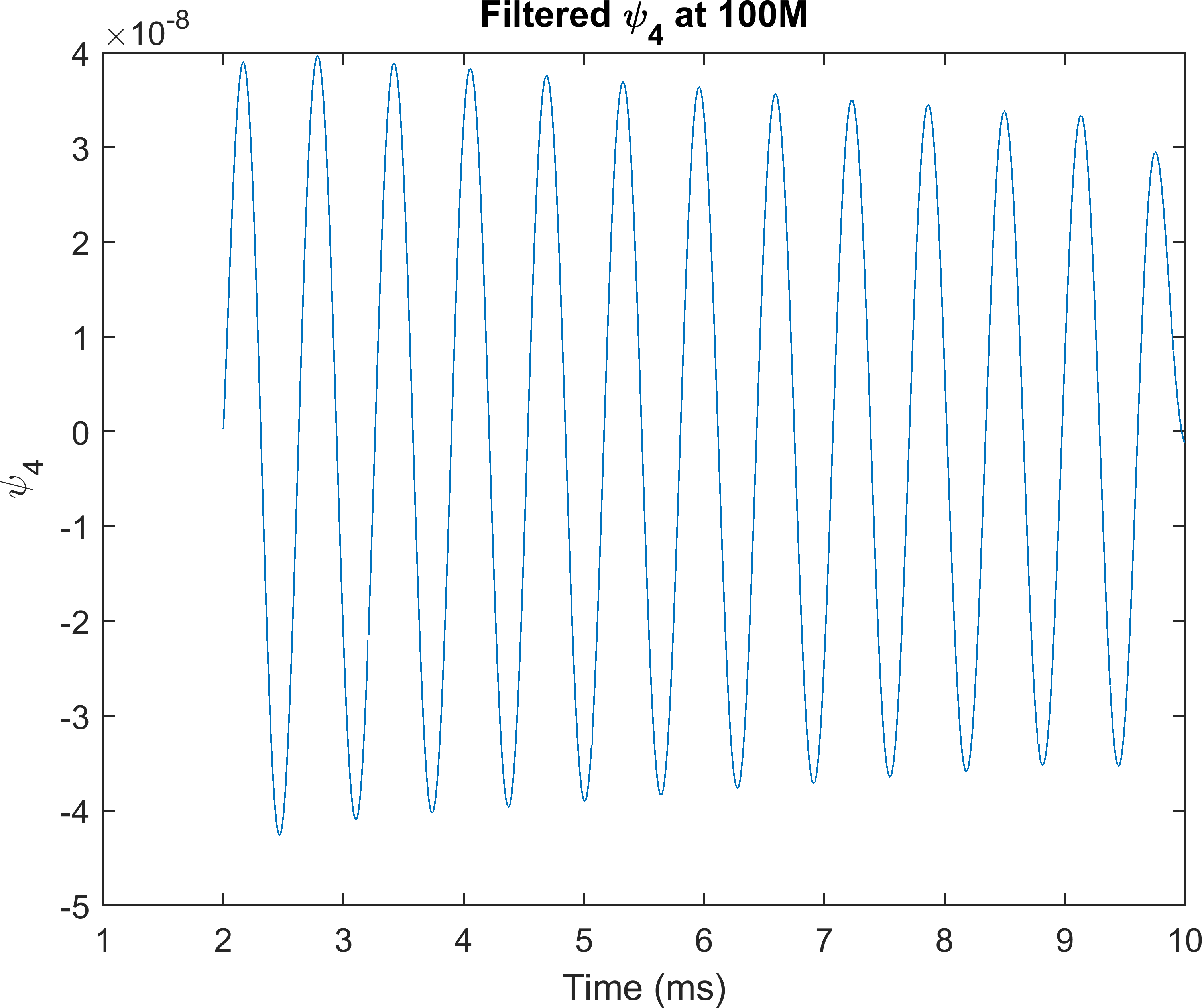}
	}
	\caption{Plot of the $\psi_4$ data extracted at 100$\text{M}$. The left plot depicts the unfiltered data. We note that the unfiltered data is much noisier than the data extracted at 10$\text{M}$ in Fig~\ref{fig:filtering_10M}, having passed through multiple refinement layers.  These refinement layers are believed to induce high frequency noise in the waveform. In the  plot, we show the post filtered data.  Compared to Fig~\ref{fig:filtering_10M}, the data has been significantly changed by the filtering.  In turn, the post filtered data appears to match the data in \ref{fig:filtering_10M} much more closely.  This supports the notion that the unfiltered $100\text{M}$ data was contaminated by high frequency noise passing through refinement boundaries.  We also note that the end of the waveform is distorted in a similar matter to Fig~\ref{fig:filtering_10M}. For this reason, we removed the end of the data from the computation of the damping times.}
	\label{fig:filtering_100M}
\end{figure*}

\begin{figure*}
\centerline{
	\includegraphics[width=0.5\linewidth]{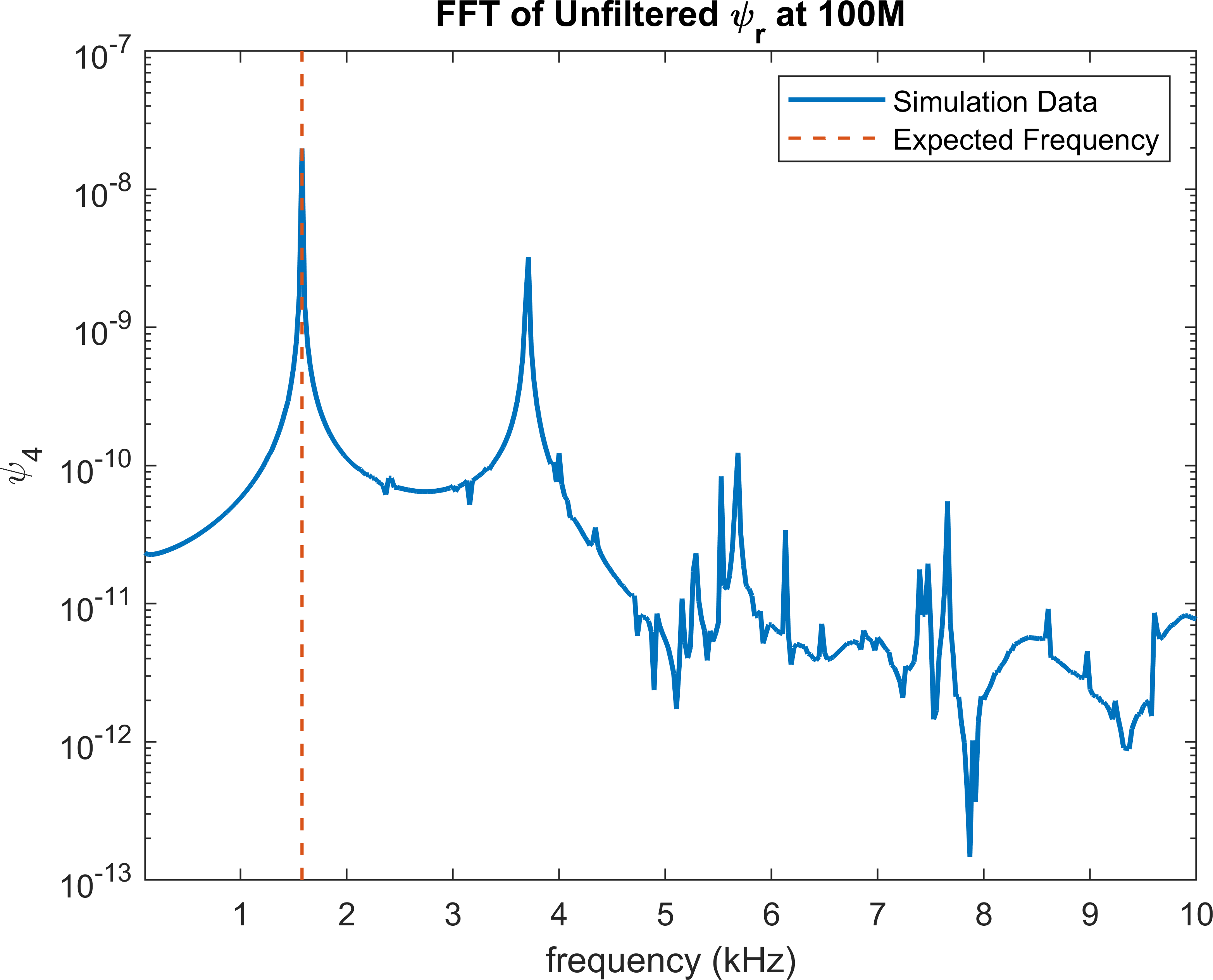}
	\includegraphics[width=0.5\linewidth]{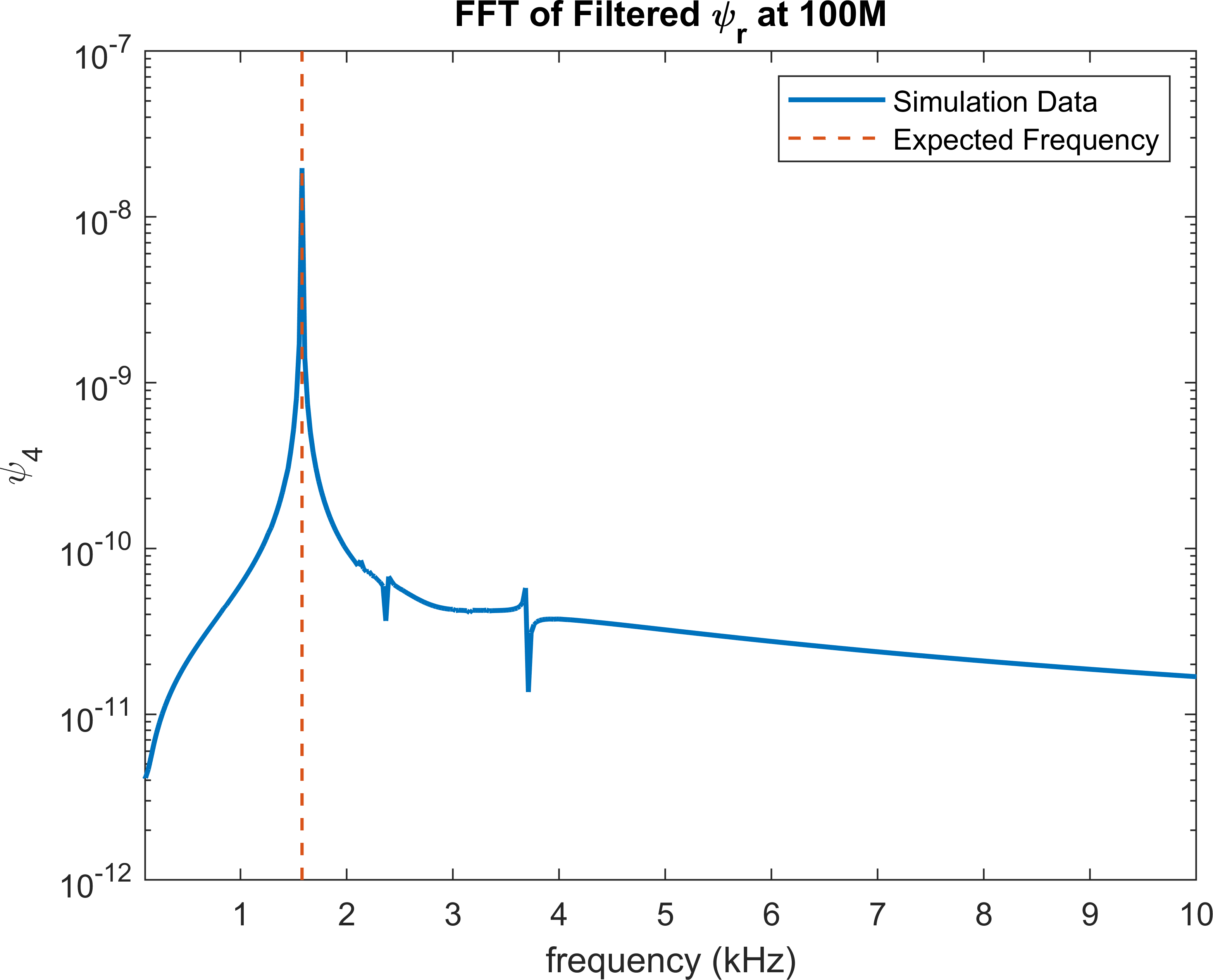}
	}
	\caption{Plots of the FFTs of the WENO HLLE simulation with a fine resolution of 0.0625$\text{M}$ before and after the filtering.  The gravitational waves here were extracted from 100M, the portion that was analyzed to extract the frequencies and damping times. The left plot illustrates the unfiltered FFT, which contains the four distinct peaks seen in \ref{fig:fft}. However, there is a significant amount of noise after the second peak.  On the right we show the FFT after the filtering has been applied.  We observed that the noise is no longer present.  However, the second peak is significantly suppressed, while the third and fourth peaks are no longer visible.  Thus we did not consider effects from these modes in our analysis of the frequency or damping times.}
	\label{fig:filtering_fft}
\end{figure*}

Initially, the $\psi_4$ data contained significant high frequency noise that interfered with our analysis.  We noted that the signals extracted from 10M, before passing through the first refinement boundary, did not exhibit any such high frequency noise. Therefore,we hypothesized that passing through the various refinement boundaries as it traveled to the relevant extraction sphere at 100M.\\
\indent 
To compensate for this high frequency behavior induced from the refinement layers, we applied a low pass filter to the raw data.  Specifically, we applied a butter filter with zero phase filtering.  Relative to the sample frequency, the filter had a passband frequency of 0.01 and a stopband frequency of 0.04.  The passband ripple amplitude of the filter was 1 dB and the stopband attenuation was 60 dB.  This filtering was essential to ensuring the extracted $\psi_4$ produced meaningful data analysis especially for the damping times. Figs \ref{fig:filtering_10M} and \ref{fig:filtering_100M} illustrates the effect of the filtering on the data.\\
\indent
We observe the frequency spectrum before and after the filtering by looking at the looking at the FFTs extracted at 100M in Fig~\ref{fig:filtering_fft}.  We observe that the unfiltered frequency spectrum is very noisy, which is most apparent after the second peak.  The filtering eliminates this noise.  However, the filtering also significantly suppresses the second peak and renders the third and fourth invisible.

\
 
\end{document}